\documentclass[aps,prl,twocolumn,groupedaddress,showpacs]{revtex4}

\usepackage{graphicx}
\usepackage{dcolumn}
\usepackage{bm}
\usepackage{amssymb}

\newcommand{\ket}[1]{\left|#1\right\rangle}

\begin{document}

\title{Dressed Relaxation and Dephasing in a Strongly Driven Two-Level system}

\author{C.M.~Wilson}\email[\textrm{e-mail:}]{chris.wilson@chalmers.se} 
\author{G.~Johansson}
\author{T.~Duty}\altaffiliation{T.\,D.'s present address: The University of Queensland, Brisbane QLD 4072 Australia.}
\author{F.~Persson, M.~Sandberg, P.~Delsing}

\affiliation{Microtechnology and Nanoscience, MC2, Chalmers University of Technology, SE-412 96
G\"oteborg, Sweden}

\date{\today}

\begin{abstract}
We study relaxation and dephasing in a strongly driven two-level system interacting with its environment.  We develop a theory which gives a straightforward physical picture of the complex dynamics of the system in terms of dressed states.  In addition to the dressing of the energy diagram, we describe the dressing of relaxation and dephasing.  We find a good quantitative agreement between the theoretical calculations and measurements of a superconducting qubit driven by an intense microwave field.   The competition of various processes leads to a rich structure in the observed behavior, including signatures of population inversion. 
\end{abstract}


\maketitle

An essential question in quantum theory is how a system is affected by its interaction with its environment.  There has been great progress in recent years describing this interaction through decoherence theory, which quantifies the effects in terms of relaxation and dephasing.  An important question in this field is how the interaction of the system with its environment is modified when the system is strongly driven.  There has been a significant theoretical effort to understand this problem \cite{GrifoniCollection}, but there remains a variety of different theoretical approaches with few experimental results to guide progress.  This has changed recently with the emergence of the new field of \textit{circuit} quantum electrodynamics \cite{DelftQED,YaleCavity}, where nanoelectronic circuits interact with photons at the quantum level.  The design flexibility afforded by these solid-state systems has allowed the exploration of new regimes of drive strength and new forms of interaction \cite{WilsonDS, MITLZ, FinnishLZ}.  In this Letter, we use the techniques of circuit QED to make a quantitative comparison between the theory and experiment of relaxation and dephasing in a strongly driven system.


We present a microscopic model of how the driven system interacts with the quantum vacuum of the environment.  We exploit a hierarchy of energy scales to develop an analytic description which gives a straightforward physical picture of this complex system in terms of dressed states.  We show that by including a minimal number of parameters, which describe the spectral density of the vacuum, we can explain a wide variety of observed effects, including population inversion in the dressed states. 

Dressed states of superconducting circuits have recently received attention in the context of quantum information \cite{NoriDS}.  In particular, theoretical work on quantum state detection, i.e., qubit readout \cite{ChalmQCTheory}, has suggested that relaxation and dephasing of the dressed states may be an important factor limiting performance \cite{DressedCQED}. 



\begin{figure*}[t]
\includegraphics[width = 2\columnwidth]{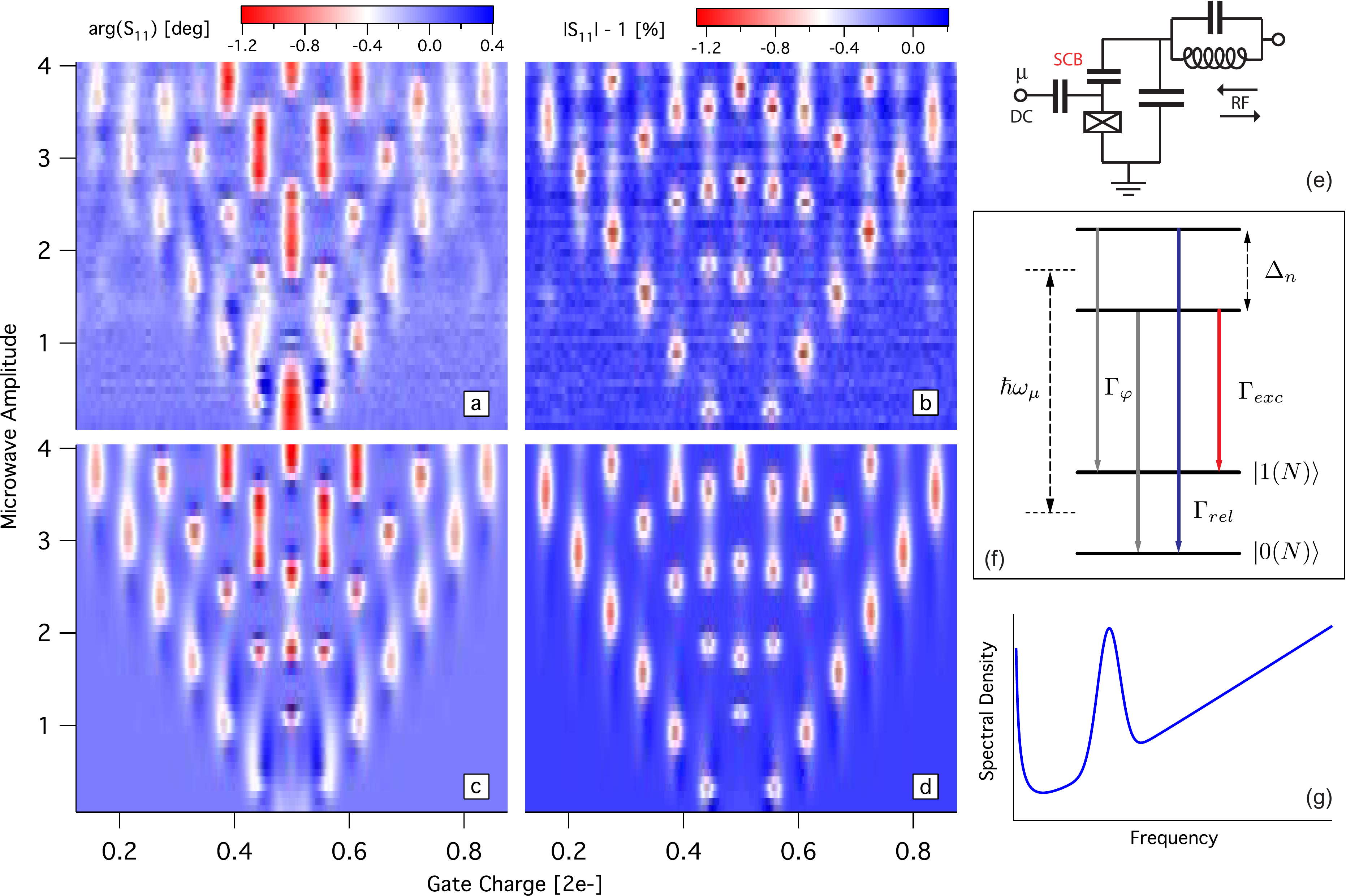}
\caption{\label{PhaseData} (Color Online) RF response of the strongly driven SCB coupled to an oscillator.  The images represent the rf reflection coefficient of the system, $S_{11}$, as a function of dc gate charge, $n_g$, and microwave amplitude, $A_{\mu}$. (a) and (b) show the data for the phase, $\arg(S_{11})$, and magnitude,  $|S_{11}|-1$, respectively.  (c) and (d) show the corresponding theoretical calculations.  The fitting parameters represent a model for the environmental charge noise, sketched in (g), which is responsible for dressed relaxation and dephasing.  The data and theory are plotted with the same color scales.  (e) Schematic of the SCB and resonator.  (f) Energy level diagram showing the dressed levels and transitions. (g) Sketch of the environmental spectral density showing the $1/f$ tail, oscillator peak, and Ohmic background.}
 \end{figure*}

Our artificial atom, the single Cooper-pair box (SCB), is composed of a superconducting Al island connected to a superconducting reservoir by a small area Josephson junction \cite{ButtikerQC}.  The two charge-basis states of the SCB represents the presence (absence) of a single extra Cooper-pair on the island.  The  Hamiltonian of the SCB coupled to the driving microwave field is $H = -\frac{1}{2}E_{Ch}\sigma_z - \frac{1}{2}E_J\sigma_x+ \hbar\omega_{\mu} a^{\dagger}a + g\sigma_z(a + a^\dagger)$, where $\sigma_i$ are the Pauli spin matrices and $a^\dagger$, $a$ are the creation and annihilation operators for the microwave field.  The first two terms represent the uncoupled SCB Hamiltonian, where $E_{Ch} = E_Q(1-2n_g)$ is the electrostatic energy difference between the ground and excited state of the qubit and $E_J$ is the Josephson coupling energy.  Here $E_Q = (2e)^2/2C_{\Sigma}$ is the Cooper-pair charging energy, $C_{\Sigma}$ is the total capacitance of the island, and $n_g = C_g V_g/2e$ is the dc gate charge used to tune the SCB.  We contact the island with a two junction SQUID which allows us to tune $E_J$ with a small magnetic field.  The third term represents the free driving field, and the last term represents the coupling,  with strength $g$, between the SCB and the microwave amplitude.


We measure the dressed states by coupling the driven SCB to an rf oscillator (Fig. \ref{PhaseData}(e)).  We probe the oscillator with a weak rf field, measuring the magnitude and phase of the the rf reflection coeffient, $S_{11}$.  In a typical measurement, we use an external magnetic field to fix a value for $E_J$.  We then produce a 2D map of $S_{11}$ by slowly sweeping $n_g$ while stepping the microwave amplitude.  For more details of the experimental setup, see \cite{WilsonDS}.  In Fig. \ref{PhaseData}(a) \& (b), we present measurements of $S_{11}$ for representative values of $E_J/h = 2.6$ GHz, $E_Q/h=62$ GHz, $\omega_{\mu}/2\pi = 7$ GHz and  $\omega_{rf}/2\pi = 0.65$ GHz.  We see a rich response in both the magnitude and phase.

To understand the data, we start by considering the Hamiltonian of our system, $H$, in a region of $n_g$ where $E_{Ch} \sim n\hbar\omega_{\mu}$. To first order, we get a ladder of energies 
 \begin{displaymath}
\label{bands}
E_{1(0)}^N = N\hbar\omega_{\mu}\pm \frac{1}{2}\sqrt{\left(E_{Ch}-n\hbar\omega_{\mu}\right)^2 + \left(\Delta_{n}(\alpha)\right)^2}
\end{displaymath}
with pairs of an "excited" state, $\ket{1(N)}$, and a "ground" state, $\ket{0(N)}$, that repeat for different photon numbers $N$ (Fig. \ref{PhaseData}(f)) \cite{Foot1, PeggSeries, NECMultiRabi}.  Here $\Delta_{n}(\alpha) = E_J J_{n}(\alpha)$ is the dressed gap between these states, which varies with the normalized microwave amplitude $\alpha =2E_Q n_{\mu}/\hbar\omega_{\mu}$, $n_{\mu}  \equiv \gamma_{\mu} A_{\mu}/2e$ being the microwave amplitude in units of $2e$.  Here $A_{\mu}$ is the microwave amplitude at the generator and $\gamma_{\mu}$ is the microwave coupling.  Finally, $J_n$ is a Bessel function of order $n$.

\begin{figure}
\includegraphics[width=1\columnwidth]{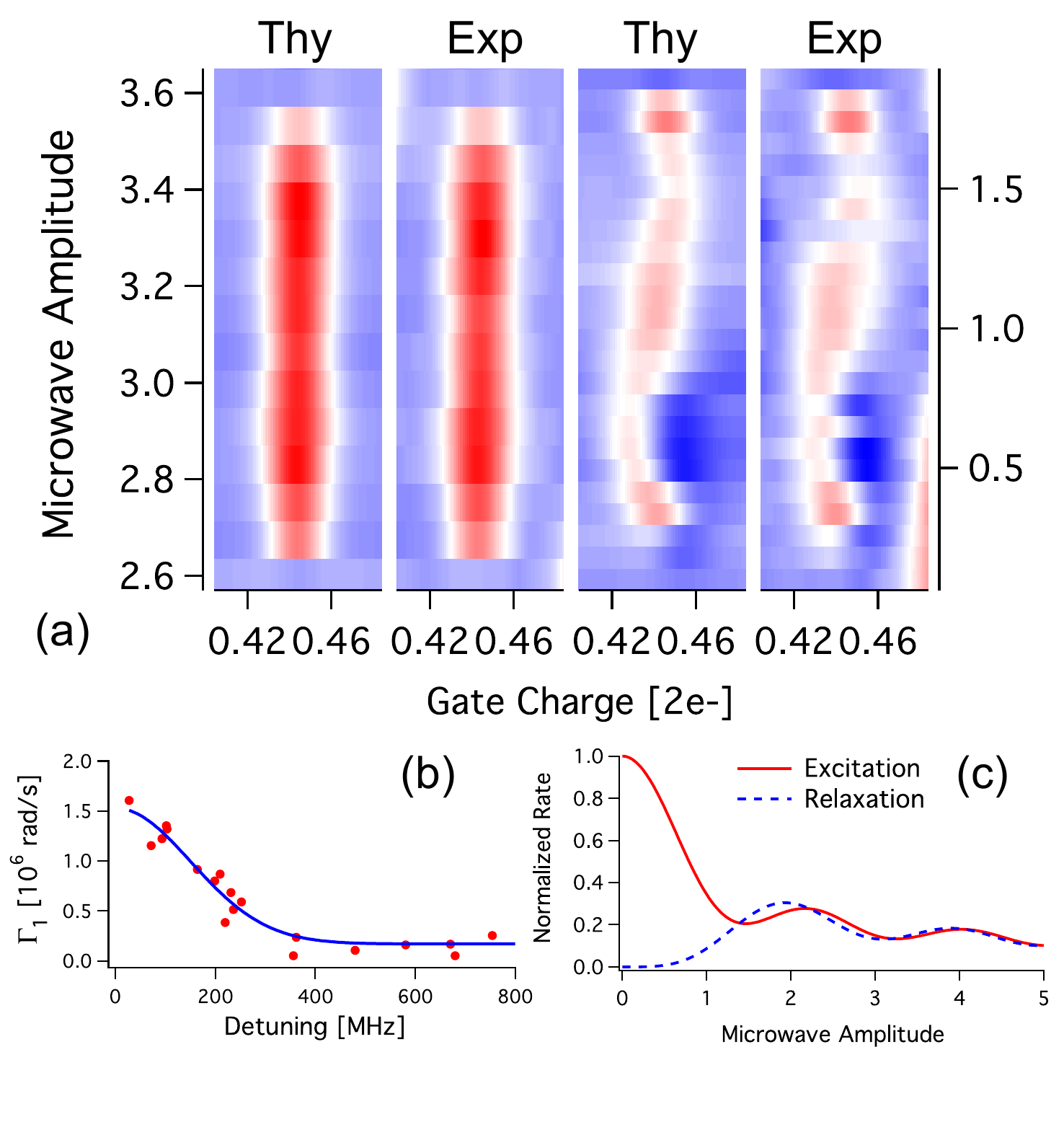}%
\caption{\label{BlochResponse}(Color Online) (a) Dressed bloch response of the 1-photon dressed state.  The (color) scale is the same as Fig. \ref{PhaseData}.  The two left panels are the theory and data for higher amplitude, while the right panels are for lower amplitude.  (b) The extracted values of $\Gamma_{rel}^{std}$ as a function of the detuning between the dressed states and oscillator, for positive detuning.  (c) The calculated values of $\Gamma_{exc}(\eta = \pi)$ and $\Gamma_{rel}(\eta = \pi)$ (excluding $\Gamma_{rel}^{std}$) as a function of the microwave amplitude $A_{\mu}$.   
}
\end{figure}

We have previously shown that the absorption features in Fig. \ref{PhaseData}(b) arise from the resonant interaction of the dressed states and the readout oscillator \cite{WilsonDS}.  However, the phase response is more varied.  Referring to Fig. \ref{PhaseData}(a), we see that there is a cone at high power where the response is relatively simple, showing vertical stripes where the phase shift is unimodal.  We showed this response could be explained well by the dispersive shift of the oscillator frequency caused by its coupling to the near resonant dressed states, an effect related to the quantum capacitance of the states \cite{SillanpaaQC, DutyQC}.  At lower powers, we see that there are regions where the phase shift becomes bimodal, changing signs as a function $n_g$ when moving across a resonance.  This is seen clearly in  Fig \ref{BlochResponse}(a) as a change in color from red to blue.  We note that the transition in the character of the response does not occur at a simple uniform threshold.

To explain this rich variety of phenomena, we must understand how relaxation and dephasing are dressed in our strongly driven system.  We start by considering transitions between states in this ladder which are induced by the system's coupling to the environmental bath, seen through the oscillator.  We do this by writing the master equation describing the dynamics of the density matrix, $\rho$, which spans our set of states.  The master equation contains a set of transitions rates which are proportional to the square modulus of the matrix elements of the operator which couples the system to the bath.  To calculate these matrix elements, we first extend the calculation of our dressed states to next order in $E_J/m\hbar\omega_{\mu}$.  We include the contribution of all off-resonant states at this order, including the effects of $m$-photon transitions for arbitrary values of $m$.  After calculating the rates, we then reduce the full master equation to a master equation for an effective two-level system by tracing over $N$.  It is this reduced master equation that we use to describe the dynamics of the reduced dressed states, $\ket{1}$ and $\ket{0}$, interacting with our readout oscillator.  

We first consider the effects of charge noise in the environment, which is typically the dominant source of noise for a SCB.  The noise couples through $\sigma_z$ and has a spectral density $S_Q(\omega)$.  We start by assuming that the bath is at zero temperature.  The reduced master equation then includes three rates: the relaxation rate, $\Gamma_{rel}(\eta) = \Gamma_{rel}^{std} + \sum_{m>0} \Gamma_{rel}^{m}$; the excitation rate, $\Gamma_{exc}(\eta) = \sum_{m>0} \Gamma_{exc}^{m}$; and the dephasing rate, $\Gamma_{\varphi}(\eta) = \Gamma_{\varphi}^{std} + \sum_{m>0} \Gamma_{\varphi}^{m}$.   Here $\eta$ is the mixing angle across the dressed degeneracy point defined by $\tan\eta = \Delta_n/(E_{Ch}-m\hbar\omega_{mu})$. The different rates are $\Gamma_{rel}^{std}  = A^2 \sin^2\eta S_Q(\Delta)$ and $ \Gamma_{\varphi}^{std} = A^2 \cos^2\eta S_Q(0) + \frac{1}{2}
\left(\Gamma_{rel}+\Gamma_{exc}\right)$ and 

\begin{eqnarray}
\label{relax}
\Gamma_{rel}^{m} = B_m
\left[\cos^2\frac{\eta}{2} J_{m-n}(\alpha) +
\sin^2{\frac{\eta}{2}} J_{-(m+n)}(\alpha) \right]^2,
\\
\label{excite}
\Gamma_{exc}^m =  B_m \left[\sin^2\frac{\eta}{2} J_{m-n}(\alpha) +
\cos^2{\frac{\eta}{2}} J_{-(m+n)}(\alpha) \right]^2,
\\
\label{dephase}
\Gamma_\varphi^m =  \frac{B_m\sin^2\eta} {4} \left[J_{m-n}(\alpha) - J_{-(m+n)}(\alpha) \right]^2 ,
\end{eqnarray}
where $A \equiv 1- (E_J/2\hbar\omega_\mu)^2 \sum_{m\neq 0}
J_{m-n}^2(\alpha)/m^2$ and $B_m \equiv E_J^2 S_Q(m\hbar\omega_{\mu}) / ( m \hbar\omega_{\mu})^2$.  We can include the effects of a finite bath temperature, $T$, by adding to $\Gamma_{exc}$ the term $\Gamma_{rel} \exp(-\hbar\Delta/k_B T)$. A few comments are in order.  First, we note that $\Gamma_{rel}^{std}$ and $ \Gamma_{\varphi}^{std}$ are the standard rates we would calculate for an undriven two-level system coupled to a bath (assuming $A \sim 1$).  These rates represent transitions that do not change $N$.  The other rates, $\Gamma^m$, represent transition that change $N$ by $m$ photons.  Next, we note that we have an effective excitation rate even though we have only included the effects of charge relaxation in the full master equation.  This is related to transitions that take us from a ground dressed state, $\ket{0(N)}$, to an excited dressed state with fewer photons, $\ket{1(N-m)}$ (Fig.\ref{PhaseData}(f)).  In this transition, the total system loses energy to the bath, but our effective two-level system appears to be excited.

We model the readout of the dressed states in the following way.  Starting from the reduced master equation, we derive the semiclassical Bloch equations for the dressed state charge driven by the small rf probe voltage $V_{rf} = 2en_{rf}/C_{rf}$.  Besides the three rates above, the Bloch equations also include an effective temperature, $T_{eff}$, which determines the equilibrium occupation of the states.  This is \textit{calculated} as a function of $\alpha$ and $\eta$ as 
\begin{displaymath}
T_{eff} = \frac{\Delta_n(\alpha)}{k_B \ln\left( \frac{\Gamma_{rel}(\eta)}{\Gamma_{exc}(\eta)}\right)}.
\end{displaymath}
We then take the standard solution of the Bloch equations for the in-phase and quadrature component of the dressed state charge, $Q_i$ and $Q_q$.  These are used to calculate an effective parallel resistance and capacitance, $R_{eff} =  V_{rf}/\beta\dot{Q_q} = V_{rf}/2\pi f_c \beta Q_q$ and $C_{eff} = \beta Q_i/V_{rf}$, which are then used to calculate the reflection coefficient $S_{11}$.  This is essentially the same procedure used in \cite{WilsonDS} except that we now include the calculated $T_{eff}$.  In addition, the coupling constant $\beta = C_{rf}/C_{\Sigma}$ has been added to better account for the coupling of the dressed state charge to the oscillator.

To compare our theory with experiment, we need to have a model for the environmental spectral density $S_Q(\omega)$ (see Fig.\ref{PhaseData}g).  The typical starting point is to assume that $S_Q(\omega)$ is Ohmic with an additional contribution from $1/f$ charge noise \cite{MakhlinReview}.  Also, because of the readout oscillator, $S_Q(\omega)$ will not be smooth around $\omega_{rf}$.  For the purposes of fitting, we will then describe the environment by three parameters: $S_Q(\omega \approx 0)$ which determines $\Gamma_{\varphi}^{std}$, $S_Q(\omega = \Delta_n)$ which determines $\Gamma_{rel}^{std}$, and a high-frequency coupling constant $\kappa$, such that $S_Q(\omega >> \Delta_n) \varpropto \kappa^2\omega R_0$.  $\kappa$ then determines the rates for the $m$-photon relaxation processes.  We note that compared to the simple model in \cite{WilsonDS}, we have only added one fit parameter, which is $\kappa$.
 
 In Fig. \ref{BlochResponse}(a), we show the results of performing a detailed fit to the one-photon resonance of the data in Fig.  \ref{PhaseData}.    The magnitude and phase data are fit simultaneously, which is important for the parameters to be constrained, although we only show the phase data.  For compactness, the vertical axis is split and the higher and lower lobes are plotted side-by-side.  We see that the agreement between data and theory is very good for both lobes of the response, despite the fact that the experimental responses look very different.  The three parameters mentioned above are allowed to vary independently for each value of $\alpha$.  However, we find that the variation of the value of $\kappa$ is less than 10\%, which is comparable to the random error in each fit.  The extracted value of $S_Q(\hbar\omega_{\mu})$ would translate into a relaxation time for a typical charge qubit of $1/\Gamma_{rel} \sim 300$ ns, consistent with observed values. In Fig. \ref{BlochResponse}(b), we plot the extracted values of the $\Gamma_{rel}^{std}$, proportional to $S_Q(\Delta_n)$.  We see that $S_Q(\omega)$ is peaked around the oscillator frequency $\omega_{rf}$, as we expect and as we observed before \cite{WilsonDS}.  It is worth noting again, that despite the very different appearance of the response in the two lobes, the extracted rates are consistent with each other, falling on the same curve.  Taken together, these results confirm that we have a good understanding of how charge relaxation takes place in the extended ladder of dressed states, and how that reduces to relaxation and excitation in the reduced basis $\ket{1}$ and $\ket{0}$.

It is possible to understand the response in straightforward physical terms within our dressed state interpretation.  First of all, we can understand the bimodal nature of the phase response at lower $\alpha$ (right side of Fig. \ref{BlochResponse}(a)).  If we consider the $n$-photon resonance on one side of the dressed degeneracy, near $\eta \sim 0$, the dominate terms in the rates (eqs. \ref{relax}-\ref{excite}) are $\Gamma_{rel}(\eta) \varpropto J_0^2(\alpha) \sim (1 - \alpha)^2$ and $\Gamma_{exc}(\eta) \varpropto J_{n+1}^2(\alpha) \sim \alpha^{2(n+1)}$.  Clearly then, for $\alpha << 1$ relaxation dominates.  On the other side, at $\eta \sim \pi$, we find instead $\Gamma_{exc}(\eta) \varpropto J_0^2(\alpha)$ and $\Gamma_{rel}(\eta) \varpropto J_{n+1}^2(\alpha)$, implying that excitation is in fact dominant.  This implies that the reduced states $\ket{1}$ and $\ket{0}$ will become inverted in this region.  At the degeneracy point, $\Gamma_{rel}(\eta) \approx \Gamma_{exc}(\eta)$ (if we ignore the standard relaxation for now) and we expect the populations to equalize.  The Bloch equations tell us that the phase response of the excited state has the opposite sign from the ground state response.  Therefore, as we move across the dressed degeneracy point, we expect the phase shift to start with one value, go to zero, and then change sign.  This is exactly the bimodal shape that we observe.  

As we move to higher values of $\alpha$, the situation changes.  The dominant $J_0$ term in the rates decays sharply, leaving the summed contributions from many different photon transitions.  As shown in Fig. \ref{BlochResponse}(c), these summed rates become essentially equal for larger amplitudes.  Essentially, the strongly driven transition saturates.  Once this happens, the dynamics of the reduced dressed states are determined only by $\Gamma_{rel}^{std}$.  This is why we see a crossover to the simple form of the response at higher $\alpha$.  

Having shown that we can fit the response for particular values of $\alpha$, we can also wonder if it is possible to explain the response over the whole range of parameters.  To do this, we need to account for additional dephasing which is not captured by (\ref{dephase}).  This is not in fact surprising since the first order contribution should vanish at the dressed degeneracy points.  We use an adiabatic approximation to model the second order charge noise through a pure dephasing term connected to an effective $\sigma_x$ operator \cite{Makhlin2ndOrder}.  To lowest order in $E_J/m\hbar\omega_{\mu}$, we find that the contribution of $m$-photon transitions is
\begin{equation}
\label{DephaseX}
\Gamma_{\varphi,x}^{m}(\eta) =  \sin^2\eta \left( J_{-(m+n)}+J_{m-n}\right)^2 S_X(m\hbar\omega_\mu) 
\end{equation}
where $S_X(\omega)$ is then the spectral density of the noise associated with the effective $\sigma_x$ operator.  


In Fig. \ref{PhaseData}(c) \& (d), we show calculations of the response covering the full range of data.  We no longer fit the parameters associated with relaxation in these calculations.  Instead, we model the environment based on the parameters extracted from the fits presented in Fig. \ref{BlochResponse}.  In modeling the additional dephasing, we have assumed that the spectrum has an effective cutoff above $\hbar\omega_\mu$, i.e. we only include the $m=0,1$ transitions.  This is justified by the adiabatic approximation mentioned above. Therefore, there is a total of just 2 fitting parameters for both 2D plots combined, $S_X(0)$ and $S_X(\hbar\omega_\mu)$.  We see that the agreement for the magnitude data is striking and that it is also quite good for the phase data.  In particular, we reproduce both the quantitative level of the response and all the qualitative features mentioned earlier.  (The possible exception to this statement is the 0-photon phase response at small $A_{\mu}$.  This is not surprising, however, as in this region the dressed gap is far detuned from the resonator and our model for the response does not apply.) Taken as a whole, this is strong evidence for our interpretation.

We thank Sartoon Fattahpoor for useful discussions. This work was supported by the Swedish VR and SSF,  the Wallenberg foundation, and by the European Union through the project EuroSQIP.  The samples were made at the nanofabrication laboratory at Chalmers.


\begin{thebibliography}{99}
\bibitem{Foot1}
Eqn. (2) of Ref. \cite{WilsonDS} should be divided by 2.
\end{thebibliography}
\end{document}